\documentclass[prd,amsmath,amssymb,superscriptaddress,floatfix,nofootinbib,10pt]{revtex4}
\usepackage{times}
\usepackage{amssymb,amsbsy,amsmath,amsfonts}
\usepackage{graphicx}
\usepackage{float}
\usepackage{color}
\usepackage{morefloats}
\usepackage{rotating}
\usepackage{srcltx}
\usepackage{slashed}
\usepackage{multirow}
\usepackage{verbatim}
\usepackage{hyperref}
\usepackage{tabularx}
\usepackage{bm}
\allowdisplaybreaks[4]

\usepackage{bbding}
\usepackage{threeparttable}

\usepackage{mathrsfs}

\DeclareUnicodeCharacter{2212}{\ensuremath{-}}

\newcommand{\PreserveBackslash}[1]{\let\temp=\\#1\let\\=\temp}
\newcolumntype{C}[1]{>{\PreserveBackslash\centering}p{#1}}
\newcolumntype{R}[1]{>{\PreserveBackslash\raggedleft}p{#1}}
\newcolumntype{L}[1]{>{\PreserveBackslash\raggedright}p{#1}}

\begin{document}

\title{The  $\Lambda_c^+\to\Lambda\pi^+\pi^+\pi^-$ reaction, and a triangle singularity producing  the $\Sigma^*(1430)$ state }

\author{Yi-Yao Li}
\affiliation{
State Key Laboratory of Nuclear Physics and 
Technology, Institute of Quantum Matter, South China Normal 
University, Guangzhou 510006, China}
\affiliation{Key Laboratory of Atomic and Subatomic Structure and Quantum Control (MOE), Guangdong-Hong Kong Joint Laboratory of Quantum Matter, Guangzhou 510006, China }
\affiliation{ Guangdong Basic Research Center of Excellence for Structure and Fundamental Interactions of Matter, Guangdong Provincial Key Laboratory of Nuclear Science, Guangzhou 510006, China  }
\affiliation{Departamento de Física Teórica and IFIC, Centro Mixto Universidad de Valencia-CSIC Institutos de Investigación de Paterna, 46071 Valencia, Spain}

\author{Jing Song}
\email[]{Song-Jing@buaa.edu.cn}
\affiliation{School of Physics, Beihang University, Beijing, 102206, China}%
\affiliation{Departamento de Física Teórica and IFIC, Centro Mixto Universidad de Valencia-CSIC Institutos de Investigación de Paterna, 46071 Valencia, Spain}

\author{ Eulogio Oset}
\email[]{oset@ific.uv.es}
\affiliation{Departamento de Física Teórica and IFIC, Centro Mixto Universidad de Valencia-CSIC Institutos de Investigación de Paterna, 46071 Valencia, Spain}
\affiliation{Department of Physics, Guangxi Normal University, Guilin 541004, China}

\author{Wei-Hong Liang}
\email[]{liangwh@gxnu.edu.cn}
\affiliation{Department of Physics, Guangxi Normal University, Guilin 541004, China}
\affiliation{Guangxi Key Laboratory of Nuclear Physics and Technology,
Guangxi Normal University, Guilin 541004, China}

\author{Raquel Molina}
\email[]{raquel.molina@ific.uv.es}
\affiliation{Department of Physics, Guangxi Normal University, Guilin 541004, China}
\affiliation{Departamento de Física Teórica and IFIC, Centro Mixto Universidad de Valencia-CSIC Institutos de Investigación de Paterna, 46071 Valencia, Spain}

\begin{abstract}
We study the decay $\Lambda_c^+ \to \Lambda \pi^+ \pi^+ \pi^-$, focusing on the production of the $\Sigma^*(1430)$ resonance observed by the Belle Collaboration. Interpreted as a dynamically generated state from meson-baryon interactions in the chiral unitary approach, the $\Sigma^*(1430)$ signal is shown to be enhanced by a triangle singularity involving intermediate $K^{*-}$, $p$, and $\bar K^0$ states. This mechanism leads to a sharp peak near 1434 MeV in the $\pi^+ \Lambda$ invariant mass distribution, in agreement with the experimental observations,  and predicts a secondary peak around 1875 MeV in the $\pi^- \Sigma^*(1430)$ spectrum tied to the triangle singularity. We also estimate the branching ratio of $\Lambda_c^+ \to \pi^+ \pi^- \Sigma^*(1430)$ to be about $3.5 \times 10^{-4}$. The results for the branching ratio and the  $\pi^- \Sigma^*(1430)$  mass distributions are predictions of the theoretical approach, which could be tested with reanalysis of existing data.

\end{abstract}

\maketitle
\section{\large\bfseries Introduction}
\label{intro}
The Belle collaboration recently found a clear signal of the $\Sigma^*(1430)$ state in the decay process $\Lambda_c^+ \to \Lambda \pi^+ \pi^+ \pi^-$ \cite{Belle:2022ywa}.
The $\Sigma^*(1430)$ is a low-lying excited state of the $\Sigma$ baryon family that is not yet firmly established in the Particle Data Group (PDG) listings. However, increasing theoretical and experimental efforts suggest its relevance in processes involving $\Lambda \pi$ interactions in the $I=1$ channel. Its proximity to the $\bar{K}N$ threshold and narrow width makes it a good candidate to be dynamically generated from meson-baryon interactions, rather than a conventional three-quark excitation \cite{Oller:2000fj, Hyodo:2011ur, Roca:2013cca, Jido:2003cb, Oset:1997it, Inoue:2001ip, Garcia-Recio:2002yxy}.
This state was already predicted in earlier theoretical studies \cite{Oller:2000fj, Jido:2003cb} using the chiral unitary approach in coupled channels. In Ref.~\cite{Oller:2000fj} it appeared as a slightly bound state, while in Ref.~\cite{Jido:2003cb} it was slightly unbound, producing a cusp effect.
This cusp is also produced in a recent analyzing from LQCD data with the WT (Weinberg-Tomozawa)  interaction term, while a resonance is generated at NLO~\cite{Zhuang:2024udv}.
In Ref.~\cite{Roca:2013cca} the authors analyzed the photoproduction data of the $\gamma  p \to K^+\pi^\pm\Sigma^\mp$ of the CLAS experiment~\cite{CLAS:2013rjt} and also concluded the existence of an $I=1$ state at the threshold of  $\bar K N$, which appeared as a cusp.
These results encourage us to study hadronic decays of heavy baryons in search of that state.
The recent observation of that state in the $\Lambda \pi^+$ and $\Lambda \pi^-$ invariant mass distributions of the  $\Lambda_c^+ \to \Lambda \pi^+ \pi^+ \pi^-$   reaction, offers us a unique opportunity to dig into the structure of this state and the mechanism by which it is produced in this reaction.

\begin{figure}[H]
\begin{center}
\includegraphics[width=0.5\textwidth]{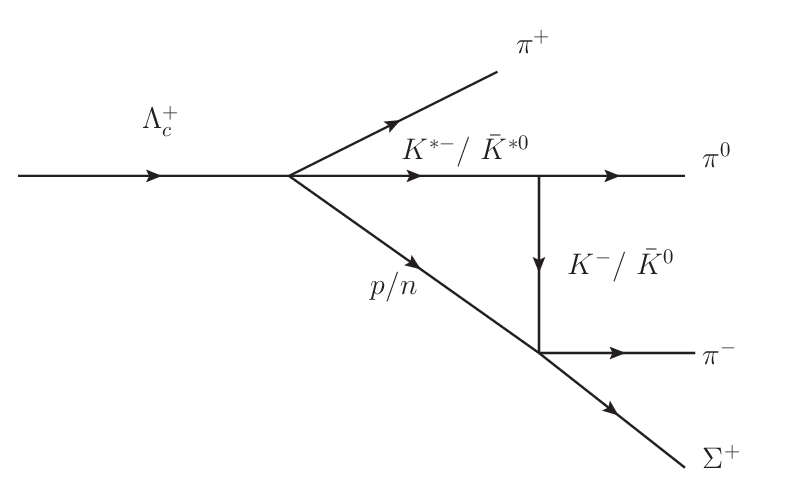}
\end{center}
\vspace{-0.7cm}
\caption{Triangle diagrams for the decay of $\Lambda_c^+ \to \pi^+
\pi^0 \pi^-\Sigma^+$ with the final state interaction of
$\bar{K}N \to \pi^- \Sigma^+$.%
\label{Fig:intro}}
\end{figure}

Given the fact that the meson-meson amplitudes for the $\Sigma^*(1430)$   state are small~\cite{Roca:2013cca}, it was suggested that using reactions in which the resonance would be produced via a triangle singularity mechanism the signal for the production of the resonance would be appreciably enhanced, giving more chances for the observation of the state. In this line, in Ref.~\cite{Xie:2018gbi} the  $\Lambda_c^+ \to \pi^+ \pi^0 \pi^-\Sigma $   reaction was proposed to look for the $\Sigma^*(1430)$  state via the mechanism depicted in Fig.~\ref{Fig:intro}, which produces the $\Sigma^*(1430)$  state from  $K^-p~(\bar K^0 n) \to \pi^-\Sigma^+$    and develops a triangle singularity for $M_\text{inv}(\pi^0 \pi^-\Sigma^+)\sim1875~\mathrm{MeV}$.
In the conclusions of that work it was also suggested to look for the   $\Sigma^*(1430)$ state in the  $\Lambda \pi^+$,  $\Lambda \pi^-$    invariant mass distributions of the $\Lambda_c^+ \to \Lambda \pi^+ \pi^+ \pi^-$ decay. This is indeed the experiment of Belle, which has been successful, showing a clear signature for the state around the $\bar K N$  threshold, as predicted.

Triangle singularities (TS) are a powerful mechanism to explain anomalous  peaks in hadronic invariant mass distributions.
They happen in reaction mechanisms proceeding via a Feynman triangle  diagram when all particles in the loop diagram can be on-shell simultaneously, and they are collinear, plus the mechanism can proceed at the classical level (Coleman-Norton theorem)~\cite{Coleman:1965xm}, leading to a kinematic enhancement that can look like a  real resonance \cite{Landau:1959fi,Coleman:1965xm,Bayar:2016ftu,Liu:2015taa,Xie:2016lvs,Guo:2017jvc,Guo:2014qra,Jing:2019cbw,Guo:2019qcn}. 

 The theoretical framework employed to get the   $\Sigma^*(1430)$  state is  based on chiral unitary approach with coupled-channels dynamics,  where some resonances appear naturally from the hadron interactions \cite{Oller:1998zr, Oset:1997it,Kaiser:1995eg,Lutz:2001yb,Sarkar:2004jh,Hyodo:2008xr,Ramos:2016odk}. In this way,  the $\Sigma^*(1430)$ has been studied as a dynamically generated state from $\pi\Lambda$,  $\bar{K}N$,  $\pi\Sigma$,  and $\eta\Sigma$   interactions in $S$-wave, providing a natural explanation for its narrow width and mass  near 1430 MeV \cite{Khemchandani:2018amu,Oller:2000fj,Jido:2003cb,Roca:2013cca, Cieply:2016jby}.
 
Several recent studies have shown that TS mechanisms can significantly change invariant mass spectra and even simulate resonance-like enhancements that can be misidentified as new particles \cite{Aceti:2015zva, COMPASS:2020yhb,Szczepaniak:2015eza,Guo:2014qra,Guo:2019twa,Bayar:2016ftu}. Similar TS effects have been found and well explained in processes  such as $B \to D \pi \pi$ or $\Lambda_b \to J/\psi p K^-$  \cite{Liu:2019zoy, Sakai:2017hpg,Debastiani:2017dlz}.
Interesting is the effect of a triangle singularity  to determine the $\gamma p \to K^+\Lambda(1405)$  {photoproduction cross section at forward angles and small momentum transfers~\cite{Wang:2016dtb,BGOOD:2021sog}}.
Other works proposing the use of triangle mechanisms that produce some resonances to enhance its production via the appearance of a triangle singularity can be seen in Refs.~\cite{He:2025vij,Dai:2018hqb,Jing:2019cbw,Molina:2020kyu,Ikeno:2021frl,Feijoo:2021jtr,Guo:2019qcn,Huang:2024oai,Wang:2024jyk}.

From the theoretical point, there is only one work tying to understand the
Belle spectrum and the 
$\Sigma^*(1430)$ production~\cite{He:2025vij}.
In Ref.~\cite{He:2025vij}, the whole spectrum is analyzed theoretically by including the 
$\Sigma^*(1385)~(3/2^+)$ observed in the Belle experiment, the hypothetical 
$\Sigma^*(1380)~(1/2^-)$, and the $\Sigma^*(1430)$ ($\Sigma^*(1435)$ in Ref.~\cite{He:2025vij}), which is also observed by a 
smaller peak at the $\bar{K}N$ threshold. The work uses the method of 
effective Lagrangians, and the whole spectrum is reproduced by fitting 
about 19 parameters, a similar number as also used in the Belle 
analysis. Two methods are considered to accommodate the $\Sigma^*(1430)~(1/2^-)$: 
one parameterizing it as a Breit-Wigner, and the other one assuming 
that this state is dynamically generated from coupled channels as found 
in Refs.~\cite{Oller:2000fj,Jido:2003cb}. The cusp structure of the $\pi^+ \Lambda$ peak is shown to be 
better reproduced if the $\Sigma^*(1430)~(1/2^-)$ is dynamically generated.

Our aim is different. Our goal is not only to calculate the enhancement in the $\Lambda \pi^+$ invariant mass spectrum but also to compute the branching ratio for the $\Lambda_c^+ \to  \pi^+ \pi^-\Sigma^*(1430)$ decay and compare it with experimental data to further test the role of the triangle singularity mechanism in this process.
We do not aim at reproducing the Belle 
spectrum, but concentrate only on the $\Sigma^*(1430)~(1/2^-)$ peak, which we 
produce via a triangle mechanism that develops a triangle singularity, 
enhancing the $\Sigma^*(1430)~(1/2^-)$ production despite the small 
amplitudes for this resonance stemming from the chiral unitary approach. 
The interesting aspect of our approach is that we can determine, not 
only the shape of the production cross section in this reaction, but 
also the branching ratio for the $\Lambda_c^+ \to \pi^+ \pi^- \Sigma^*(1430)$ decay 
without any free parameters. On the other hand, due to the structure of 
the triangle singularity, we predict a particular shape for the $\pi^- 
\Sigma^*(1430)$ mass distribution, showing a peak around 1875 MeV, which 
could be tested through reanalysis of the Belle data. Observation of this 
peak and the measurement of the branching ratio for this decay mode 
would shed valuable light on the mechanism for the reaction and the 
nature of the $\Sigma^*(1430)$ state.

\section{\large\bfseries Theoretical Formalism}
\label{sec:formalism}
\subsection{\large\bfseries Weak Decay Mechanism at Quark Level}

\begin{figure}[H]
    \centering
    \includegraphics[width=0.55\textwidth]{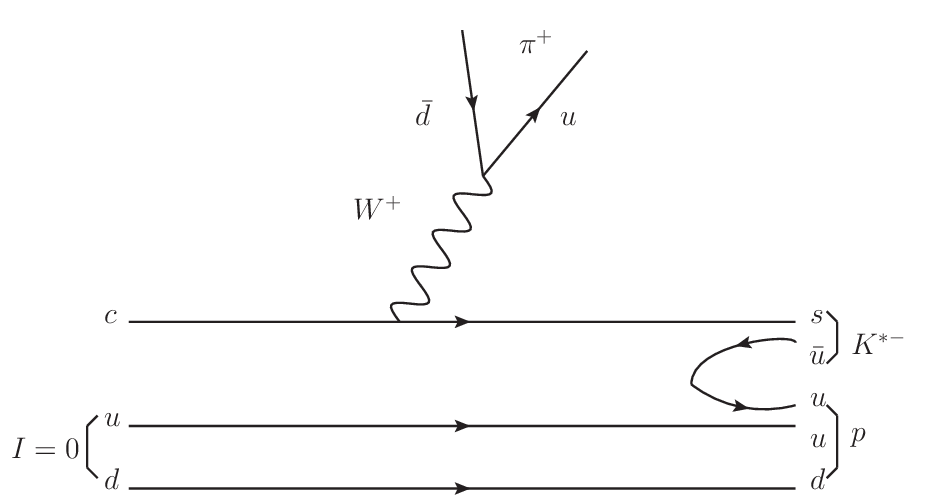}
    \caption{    External $W^+$ emission mechanism for the $\Lambda_c^+ \to \pi^+ {K}^{*-} p$ decay.}
    \label{fig:feyndiag_quark}
\end{figure}

We follow the strategy of Ref.~\cite{Xie:2018gbi} and look at the diagram of Fig.~\ref{Fig:intro}, allowing  $\Lambda_c^+ \to \pi^+ K^{*-}p$   decay. This reaction proceeds via an external $W^+$ emission mechanism as illustrated in Fig.~\ref{fig:feyndiag_quark}. 
\noindent This weak decay is Cabibbo favored, and the original $ud$ quarks inside the $\Lambda_c^+$ are spectators, retaining their isospin $I=0$ character throughout the reaction. Consequently, the $sud$ cluster after the weak vertex also has $I=0$. Hadronization occurs by inserting a $\bar{q}q$ pair with vacuum quantum numbers, which leads to the production of meson-baryon pairs. For the process of interest, the hadronization leads  to the generation of $\bar{K}^* N$ pairs in an $S$-wave configuration. Since the $ud$ pair acts as a spectator with positive parity, 
and $K^{*-}p$ in $S$-wave has negative parity, the $s$-quark should be produced with negative parity. This is an $L=1$   state. Since finally that quark is in its ground state in the $K^{*-}$, then the $s$-quark must be actively involved in the hadronization. This is depicted in Fig.~\ref{fig:feyndiag_quark}.

At the quark level, in the hadronization mechanisms using quark diagrams \cite{Miyahara:2015cja}, the $\Lambda_c^+$ decay proceeds via an external $W^+$ emission that generates a $u\bar{d}$ pair forming the $\pi^+$, while the remaining $sud$ quarks hadronize with a $\bar{q}q$ pair from the vacuum as shown in Fig.~\ref{fig:feyndiag_quark}. This hadronization can be expressed as
\begin{equation}
H = \sum_{i=1}^{3} s \bar{q}_i q_i\, \frac{1}{\sqrt{2}} (ud - du) = \sum_{i=1}^{3} M_{3i} q_i\, \frac{1}{\sqrt{2}} (ud - du),
\end{equation}
where $M_{ij}$ is the $q\bar{q}$ matrix,
\begin{equation}
M = \begin{pmatrix}
u\bar{u} & u\bar{d} & u\bar{s} \\
d\bar{u} & d\bar{d} & d\bar{s} \\
s\bar{u} & s\bar{d} & s\bar{s}
\end{pmatrix}.
\end{equation}

To obtain vector mesons in the hadronization, we rewrite the matrix $M$ in terms of the physical vectors as
\begin{equation}\label{pmatrix}
V = \begin{pmatrix}
\frac{\rho^0}{\sqrt{2}} +\, \frac{\omega}{\sqrt{2}} & \rho^+ & K^{*+} \\
\rho^- & -\frac{\rho^0}{\sqrt{2}} +\, \frac{\omega}{\sqrt{2}} & K^{*0} \\
K^{*-} & \bar{K}^{*0} & \phi
\end{pmatrix}.
\end{equation}
From this, by looking at the expressions of the baryons in terms of quarks in Ref.~\cite{Miyahara:2016yyh}, the flavor structure of the hadronized meson-baryon system $H$ can be expressed as~\cite{Dai:2018hqb}:
\begin{equation}
H = K^{*-} p + \bar{K}^{*0} n -\, \frac{\sqrt{6}}{3} \phi \Lambda,
\label{eq:H}
\end{equation}
where the $\phi \Lambda$ component is neglected due to its irrelevance in the triangle singularity (TS) mechanism considered here. The combination $(\bar{K}^{*0}, -K^{*-})$ forms an isospin doublet, thus $H$ has total isospin $I=0$, consistent with the initial $s(ud)_{I=0}$ cluster.

\subsection{\large\bfseries Triangle Diagram Mechanism}

Following the production of the $\pi^+ \bar{K}^* N$ system, the $\bar{K}^*$ subsequently decays into $\pi \bar{K}$. The $\bar{K} N$ pair then undergoes final state interaction (FSI) to produce the $\pi \Lambda$ system, with   $I=1$, which will lead to $\Sigma^*$  resonances,
in particular the $\Sigma^*(1430)$, which is dynamically generated by the meson-baryon interactions. The TS mechanism enhances this resonance's production, making it prominently visible in the $\pi \Lambda$ invariant mass distribution.

\begin{figure}[H]
    \centering    \includegraphics[width=0.45\textwidth]{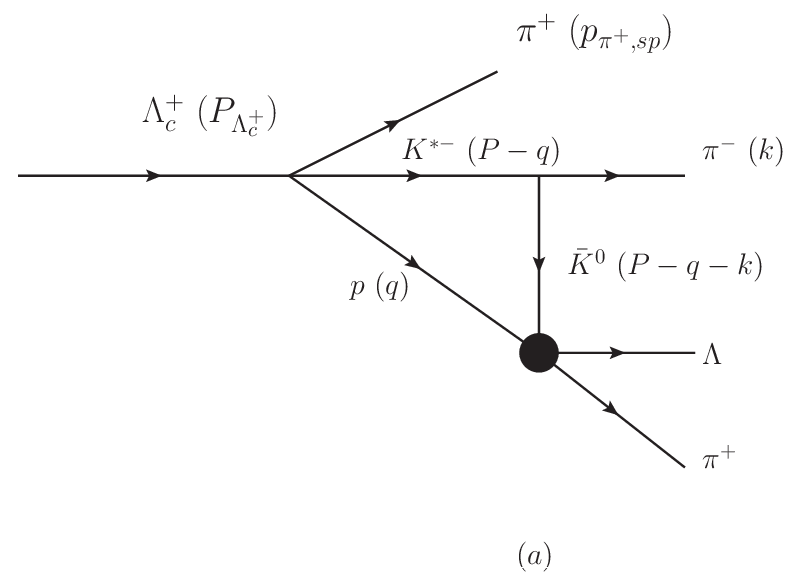}
    \includegraphics[width=0.45\textwidth]{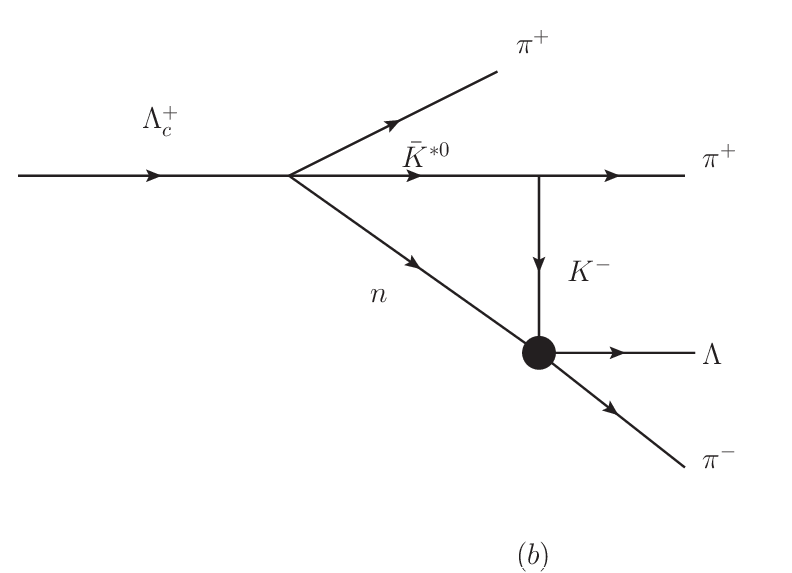}
    \caption{Triangle diagrams for $\Lambda_c^+ \to \pi^+ (\bar{K}^* N)$, followed by ${K}^{*-} \to \pi^- \bar{K}^0$ and $\bar{K}^0 p \to \pi^+ \Lambda$ (a), and $\bar{K}^{*0} \to \pi^+ {K}^-$ and ${K}^{-} n \to \pi^- \Lambda$ (b).}
    \label{fig:feyndiag}
\end{figure}

The two possible decay paths for the final state interactions are depicted in Fig.~\ref{fig:feyndiag}. 
\noindent Diagram (a) shows the $\bar{K}^0 p \to \pi^+ \Lambda$ process, while diagram (b) shows ${K}^- n \to \pi^- \Lambda$. In our calculation, we only consider the mechanism of diagram (a) which produces the  $\Sigma^*(1430)$   decaying to $\pi^+ \Lambda$.

Next, we develop the formalism for the study of the $\Lambda_c^+ \to \Lambda \pi^+ \pi^+ \pi^-$ decay, emphasizing the role of the $\Sigma^*(1430)$ resonance, which is dynamically generated in the meson-baryon interaction and appears through a TS mechanism. In particular, we analyze the decay path
\begin{equation}
\Lambda_c^+ \to \pi^+ {K}^{*-}p, \quad {K}^{*-} \to \pi^- \bar{K}^0, \quad \bar{K}^0p \to \pi^+ \Lambda,
\end{equation}
where the $\bar{K} N$ system resonates as the $\Sigma^*(1430)$ and subsequently decays into $\pi^+ \Lambda$. A triangle diagram contributes to this reaction, which can lead to a pronounced peak due to a TS under appropriate kinematic conditions.
The relevant triangle diagram, as shown in Fig.~\ref{fig:feyndiag}, proceeds via intermediate states: $\bar{K}^*$, $N$, and $\bar{K}$. By applying 
Eq.~(18)  of Ref.~\cite{Bayar:2016ftu}, it is easy to see that one should expect an enhancement of the  
 $\pi^-\Lambda\pi^+$ mass distribution around 1875~MeV. 

The decay amplitude is written as
\begin{equation}
\label{eq:triangle_amplitude}
-iT = \sum_{\text{pol of } \bar{K}^*} \int\, \frac{d^4q}{(2\pi)^4} 2M_N\, \frac{ t_{\Lambda_c^+ \to \pi^+ {K}^{*-}p}}{q^2 - M_N^2 + i\epsilon}\, \frac{ t_{{K}^{*-} \to \pi^- \bar{K}^0}}{(P - q)^2 - m_{\bar{K}^*}^2 + i\epsilon}\, \frac{ t_{\bar{K}^{0}p \to \pi^+ \Lambda}}{(P - q - k)^2 - m_{\bar{K}}^2 + i\epsilon},
\end{equation}
where $P$ is the total momentum of the ${K}^{*-}p$ system ($P=p_{\Lambda_c^+}-p_{\pi^+,sp}$),  $p_{\pi^+,sp}$ is the momentum of the spectator (sp) $\pi^+$ not involved in the loop, and $k$ is the momentum of the emitted pion from the ${K}^{*-}$ decay.

\subsection{\large\bfseries Amplitudes of \boldmath $t_{\Lambda_c^+ \to \pi^+ {K}^{*-}p}$, $t_{{K}^{*-} \to \pi^- \bar{K}^0}$, and $t_{\bar{K}^0 p \to \pi^+ \Lambda}$}


For the $\Lambda_c^+ \to \pi^+ {K}^{*-}p$    decay, we follow here Ref.~\cite{Xie:2018gbi} where the amplitude is given by
\begin{equation}
t_{\Lambda_c^+ \to \pi^+ {K}^{*-}p} = A\, \vec{\sigma} \cdot \vec{\epsilon},
\end{equation}
where $A$ is  a constant coupling strength, with  $
|A|^2 = (3.9 \pm 1.4) \times 10^{-16}~\mathrm{MeV}^{-2}$~\cite{Xie:2018gbi}, obtained from the experimental width of  the $\Lambda_c^+\to\pi^+{K}^{*-}p$ decay  in the PDG~\cite{ParticleDataGroup:2024cfk}.
The matrix $\vec{\sigma}$ acts on the spin space of the baryons, and $\vec{\epsilon}$ is the polarization vector  of the $K^{*-}$. 


The ${K}^{*-} \to \pi^- \bar{K}^0$ decay is described by the standard vector-pseudoscalar-pseudoscalar (VPP) interaction Lagrangian
\begin{equation}
\mathcal{L}_{\rm VPP} = -i g \langle [\Phi, \partial_\mu \Phi] V^\mu \rangle,
\end{equation}
where $g =\, \frac{m_V}{2f_\pi}$ with $m_V \approx 800$ MeV, and $f_\pi = 93$ MeV, $\Phi$ the $q\bar q$  matrix written in terms of pseudoscalars~\cite{Li:2024uwu} and $V$ the matrix of Eq.~(\ref{pmatrix}). This provides the amplitude
\begin{equation}
t_{\bar{K}^* \to \pi^- \bar{K}} \propto g\, \vec{\epsilon} \cdot (\vec{k}_{\bar{K}} - \vec{k}_{\pi^-}),
\end{equation}
where,
\begin{equation}
\begin{aligned}
-it_{K^{*-}\to  \pi^-\bar{K}^0}=i\mathcal{L}_{K^{*-}, \pi^-\bar{K}^0}&=gK^{*-\mu}(\pi^+\partial_\mu K^0-K^0\partial_\mu \pi^+)
\longrightarrow ig(2\vec{k}+\vec{q}) \cdot \vec{\epsilon}.\label{fomula:3}
\end{aligned}
\end{equation}


The amplitude $t_{\bar{K} N \to \pi^+ \Lambda}$ is taken from the chiral unitary approach, where the $\Sigma^*(1430)$ appears as a dynamically generated state from the $\bar{K} N$ interaction in coupled channels, as developed in Ref.~\cite{Oset:1997it}. This amplitude shows a pronounced peak around 1430 MeV. We calculate it using the framework of Ref.~\cite{Oset:1997it}, and is detailed in {subsection \ref{for_4}}.

\subsection{\large\bfseries Coalescence production}
\label{for_4}
\begin{figure}[H]
    \centering    \includegraphics[width=0.5\textwidth]{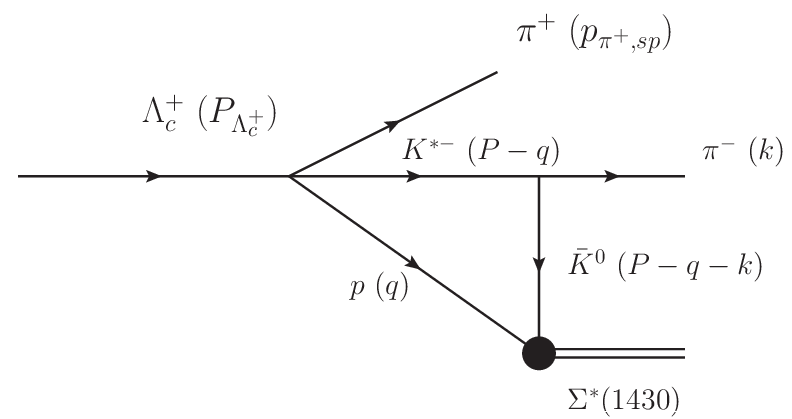}
    \caption{Coalescence triangle diagram for $\Lambda_c^+ \to \pi^+ \pi^- \Sigma^*(1430)$ decay.}
    \label{fig:1430new}
\end{figure}

Here we follow some steps which make the calculations easier. The process of {Fig.~\ref{fig:feyndiag}~(a)} has four particles in the final state which requires an elaborate phase space {evaluation} to obtain the $\Lambda_c^+$ decay width. As a first step we simplify the problem by having $\bar{K}^0p$ produce the $\Sigma^*(1430)$ resonance without worrying about its possible decay channels. The process is {called} coalescence and is depicted in {Fig.~\ref{fig:1430new}}.

Note that now we only have three particles in the final state, that can be easily handled by the formulas of the PDG~\cite{ParticleDataGroup:2024cfk}, which provide $d\Gamma/dM_{12}dM_{23}$, with $M_{12}$, $M_{23}$ two invariant masses. There are only two variables in the phase space, and $M_{13}$ can be obtained from $M_{12}$, $M_{23}$ with the  formula
\begin{equation}
M_{12}^2+M_{13}^2+M_{23}^2=M_{\Lambda_c^+}^2+m_1^2+m_2^2+m_3^2.\label{fomula:invMass}
\end{equation}
Then the triangle diagram amplitude can be evaluated by means of {Eq.~(\ref{eq:triangle_amplitude})} substituting $t_{\bar{K}^0p \to \pi^+\Lambda}\Longrightarrow g_{\Sigma^{*}(1430), \bar{K}^0p}$ with $g_{\Sigma^{*}(1430), \bar{K}^0p}$ the coupling of the $\Sigma^{*}(1430)$ to the channel $\bar{K}^0p$. Note that we are not including at this stage information about the decay of $\Sigma^{*}(1430)$ to $\pi^+\Lambda$. Only the vertex for formation from $\bar{K}^0p$ is needed at this level. Later on we shall take into account the decay to $\pi^+\Lambda$.

\subsubsection{Amplitude for $\Sigma^*(1430)$}

We denote the triangle amplitude contributing to the $\Sigma^*(1430)$ in the coalescence diagram of {Fig.~\ref{fig:1430new} } as $t_{TS}^{\Sigma^*(1430)}$. Removing the coupling of the third vertex, we write the  amplitude as:
\begin{equation}\label{fomula:Ttilde}
t_{TS}^{\Sigma^*(1430)} = g_{\Sigma^{*}(1430), \bar{K}^0p} \tilde{t}_{TS}.
\end{equation}
The calculation of the loop function of Eq.~(\ref{eq:triangle_amplitude}) is {greatly} simplified by writing a general propagator as 
\begin{equation}\label{fomula:a}
\frac{1}{q^2-m^2+i\epsilon}=\frac{1}{2\omega(\vec{q}\,)}(\frac{1}{q^0-\omega(\vec{q}\,)+i\epsilon}-\frac{1}{q^0+\omega(\vec{q}\,)-i\epsilon}),
\end{equation}
which separates the propagator into its positive energy (first term in Eq.(\ref{fomula:a})) and negative energy (second term in Eq.(\ref{fomula:a})) parts. Only for the positive energy part of the propagator, the particle can be placed on shell. Since in the TS all particles are placed on shell, we can just take the positive energy part of all the propagators. We evaluate the loop function in the rest frame of $K^{*-}p$, where $\vec{P}=0$ and then we have 
\begin{equation}
\begin{aligned}
-i\tilde{t}_{\mathrm{TS}} &= \int\, \frac{d^4q}{(2\pi)^4} A \sigma_j \epsilon_j\, \, \frac{1}{2\omega_{K^{*-}}(\vec{q}\,)}\, \, \frac{1}{P^0-q^0-\omega_{K^{*-}}(\vec{q}\,)+i\epsilon}  g(2k+q)^l\epsilon^l \\
&\quad \quad\times\, \frac{1}{2\omega_{\bar{K}^0}(\vec{q}+\vec{k})}\, \, \frac{1}{P^0-q^0-k^0-\omega_{\bar{K}^0}(\vec{q}+\vec{k})+i\epsilon}\, \, \frac{2M_N}{2E_N(\vec{q}\,)}\, \, \frac{1}{q^0-E_N(\vec{q}\,)+i\epsilon},
\end{aligned}
\end{equation}
with $\omega_i(\vec{q}\,)=\sqrt{m_i^2+\vec{q}^2},\quad E_N(\vec{q}\,)=\sqrt{M_N^2+\vec{q}^2}$. Using $\sum_{\text{pol}} \epsilon_j \epsilon_l = \delta_{jl}$, the expression becomes:
\begin{equation}
\begin{aligned}
\tilde{t}_{\mathrm{TS}} &= gA i \int\, \frac{d^4q}{(2\pi)^4}\, \frac{2M_N}{2E_N(\vec{q}\,)\, 2\omega_{K^{*-}}(\vec{q}\,)\, 2\omega_{\bar{K}^0}(\vec{q}+\vec{k})}  \sigma^j(2k+q)^j \\
&\quad \quad \times \, \, \frac{1}{P^0 - q^0 - \omega_{K^{*-}}(\vec{q}\,) + i\epsilon}\, \, \frac{1}{P^0 - q^0 - k^0 - \omega_{\bar{K}^0}(\vec{q}+\vec{k}) + i\epsilon}\, \, \frac{1}{q^0 - E_N(\vec{q}\,) + i\epsilon}.
\end{aligned}
\end{equation}
\noindent After performing the $q^0$ integration using Cauchy's theorem, we obtain:
\begin{equation}
\begin{aligned}
\tilde{t}_{\mathrm{TS}} &= gA \int\, \frac{d^3q}{(2\pi)^3}\, \frac{2M_N}{2E_N(\vec{q}\,)\, 2\omega_{K^{*-}}(\vec{q}\,)\, 2\omega_{\bar{K}^0}(\vec{q}+\vec{k})}  \sigma^j(2k+q)^j \\
&\quad \quad \times\, \, \frac{1}{P^0 - E_N(\vec{q}\,) - \omega_{K^{*-}}(\vec{q}\,) + i\epsilon}\, \, \frac{1}{P^0 - E_N(\vec{q}\,) - k^0 - \omega_{\bar{K}^0}(\vec{q}+\vec{k}) + i\epsilon}, 
\end{aligned}
\end{equation}
and using    $\int {d^3q}q^j~f(\vec{q},\vec{k})=k^j\int {d^3q}~\frac{\vec{q} \cdot \vec{k}}{|\vec{k}|^2}f(\vec{q},\vec{k})$,
\begin{equation}
\begin{aligned}
\tilde{t}_{\mathrm{TS}} &= gA \sigma^j k^j \int\, \frac{d^3q}{(2\pi)^3}\, \frac{2M_N}{2E_N(\vec{q}\,)\, 2\omega_{K^{*-}}(\vec{q}\,)\, 2\omega_{\bar{K}^0}(\vec{q}+\vec{k})} \left(2 +\, \frac{\vec{q} \cdot \vec{k}}{|\vec{k}|^2}\right) \\
&\quad \quad\times\, \frac{1}{P^0 - E_N(\vec{q}\,) - \omega_{K^{*-}}(\vec{q}\,) + i\frac{\Gamma_{K^{*-}}}{2}}\, \, \frac{1}{P^0 - E_N(\vec{q}\,) - k^0 - \omega_{\bar{K}^0}(\vec{q}+\vec{k}) + i\epsilon}  \Theta(q_{\mathrm{max}} - |\vec{q}~^*_{\mathrm{cm}}|).
\end{aligned}
\end{equation}
Note that in the last equation we have added explicitly the width of the $K^{*-}$, which is relevant as shown in \cite{Achasov:2015uua}, which is omitted in the previous formulas for simplicity. In the last equation, we have added the {factor} $\Theta(q_{\mathrm{max}}-|\vec{q}_\mathrm{cm}|)$, where $q_{\mathrm{max}}$ is the regulator of the loop function in the evaluation of the $\bar{K}N \to \pi^+\Lambda$ scattering amplitude in {Ref.~\cite{Oset:1997it}} and $\vec{q}_\mathrm{cm}$ is the momentum of the nucleon in the loop of {Fig. \ref{fig:feyndiag}} in the $\pi^+\Lambda$ rest frame. This factor {appears} because in the cut off regularization method, used in {Ref.~\cite{Oset:1997it}},  the scattering matrix appears {factorized} as $T(\vec{q},\vec{q}~')=T\Theta(q_{\mathrm{max}}-|\vec{q}~|)\Theta(q_{\mathrm{max}}-|\vec{q}~'|)$ in the rest frame of the particles ({see Ref.~\cite{Gamermann:2009uq}}). The momentum $\vec{q}$ in the loop of {Fig. \ref{fig:feyndiag}} is boosted to the $\pi^+\Lambda$ rest frame as Ref.~\cite{FernandezdeCordoba:1993az},
where
\begin{equation}\label{qcm17}
\vec{q}~^*_{\mathrm{cm}} = \left[\left(\frac{E_R(\pi^+ \Lambda)}{M_{\mathrm{inv}}(\pi^+ \Lambda)} - 1\right)\, \frac{\vec{q} \cdot \vec{k}}{|\vec{k}|^2} +\, \frac{q^0}{M_{\mathrm{inv}}(\pi^+ \Lambda)}\right]\vec{k} + \vec{q},
\end{equation}
with $E_R(\pi^+ \Lambda) \equiv \sqrt{M^2_{\mathrm{inv}}(\pi^+ \Lambda) + |\vec{k}|^2}, \quad q^0 = \sqrt{m_N^2 + |\vec{q}\,|^2}$.
We define 
\begin{equation}
\tilde{t}_{\mathrm{TS}} = \tilde{t}'_{\mathrm{TS}} \, \sigma^j k^j,
\end{equation} 
so the spin-averaged squared amplitude becomes:
\begin{equation}
\bar{\sum} \sum_{\text{pol}} |\tilde{t}_{\mathrm{TS}}|^2 = |\tilde{t}'_{\mathrm{TS}}|^2\, \, \frac{1}{2} \sum \sum_{\text{pol}} \sigma^i k^i \sigma^j k^j=|\tilde{t}~'_{\mathrm{TS}}|^2|\vec{k}|^2.
\end{equation}

\subsection{\large\bfseries $\Sigma^*(1385)$ production}

\begin{figure}[H]
    \centering    \includegraphics[width=0.5\textwidth]{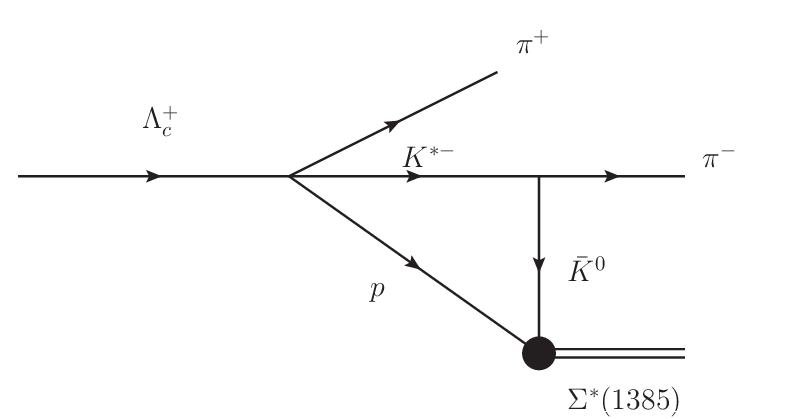}
    \caption{Coalescence production of $\Sigma^*(1385)$  through a triangle mechanism.}
    \label{fig:1385new}
\end{figure}

Next we take advantage of the structure of the triangle diagram to also produce the  $\Sigma^*(1385)~(3/2^+)$     at the coalescence level depicted in Fig.~\ref{fig:1385new}.
We anticipate here that, unlike the case of the  $\Sigma^*(1430)$  production, this mechanism does not develop a TS, as can be seen by testing Eq.~(18) of Ref.~\cite{Bayar:2016ftu}. We hence do not aim at reproducing the experimental production in the Belle experiment, but we want to see which is this contribution.

The coupling of $\bar K^0p$ to $\Sigma^*(1385)$, which is given as   $\frac{f^*}{\sqrt{3}m_\pi} \vec{S}^{\, \dag} \cdot \vec{p}_{\bar{K}^0\,\mathrm{cm}}$,
is obtained by using $SU(3)$ Clebsch–Gordan coefficients
 ~\cite{McNamee:1964xq} and relating this coupling to those of $\Delta\to \pi N$~\cite{Ikeno:2021frl}. The magnitude $\vec{p}_{\bar{K}^0\,\mathrm{cm}}$  is the $\bar K^0$ momentum in the  $\Sigma^*(1385)$  rest frame.
The total amplitude of the triangle loop contributing to the $\Sigma^*(1385)$ production is given by:
\begin{equation}
\begin{aligned}
-it_{\bar{K}^0p \to \Sigma^*(1385)}=\frac{f^*}{\sqrt{3}m_\pi} \vec{S}^{\, \dag} \cdot \vec{p}_{\bar{K}^0\,\mathrm{cm}},
\end{aligned}
\end{equation}
and then we have
\begin{equation}
\begin{aligned}
t^{\Sigma^*(1385)}_{\mathrm{TS}} &= gA \sigma^j (2k+q)^j\, \frac{f^*}{\sqrt{3}m_\pi} \vec{S}^{\, \dag} \cdot \vec{p}_{\bar{K}^0\,\mathrm{cm}}\int\, \frac{d^3q}{(2\pi)^3}\, \frac{2M_N}{2E_N(\vec{q}\,)\, 2\omega_{K^{*-}}(\vec{q}\,)\, 2\omega_{\bar{K}^0}(\vec{q}+\vec{k})} \\
&\quad \quad\times \, \frac{1}{P^0 - E_N(\vec{q}\,) - \omega_{K^{*-}}(\vec{q}\,) + i\frac{\Gamma_{K^{*-}}}{2}}\, \, \frac{1}{P^0 - E_N(\vec{q}\,) - k^0 - \omega_{\bar{K}^0}(\vec{q}+\vec{k}) + i\epsilon},
\end{aligned}
\end{equation}
where, using Eq.~(\ref{qcm17}),
\begin{equation}\label{cmq}
\begin{aligned}
-\vec{p}_{\bar{K}^0\,\mathrm{cm}} &= \vec{q}~^*_{\mathrm{cm}} = \left[\left(\frac{E_R(\pi^+ \Lambda)}{M_{\mathrm{inv}}(\pi^+ \Lambda)} - 1\right)\, \frac{\vec{q} \cdot \vec{k}}{|\vec{k}|^2} +\, \frac{q^0}{M_{\mathrm{inv}}(\pi^+ \Lambda)}\right] \vec{k} + \vec{q} \equiv a\vec{k} + \vec{q}, 
\end{aligned}
\end{equation}
with $E_R(\pi^+ \Lambda) \equiv \sqrt{M^2_{\mathrm{inv}}(\pi^+ \Lambda) + |\vec{k}|^2}, \quad q^0 = \sqrt{m_N^2 + |\vec{q}\,|^2}$, where $a$ is given by the bracket $[...]$ of Eq.~(\ref{cmq}).

The magnitude {$\vec{S}^\dag$} is the spin {transition} operator from spin 1/2 to 3/2, which has  the property $\sum_MS_i|M\rangle \langle M|S^{\dag}_j=\frac{2}{3}\delta_{ij}-\frac{i}{3}\epsilon_{ijk}\sigma_k$.
After performing the spin summation, the squared amplitude is:
\begin{equation}
\bar{\sum} \sum_{\text{pol}} |t^{\Sigma^*(1385)}_{\mathrm{TS}}|^2 =\, \frac{2}{3} |Q_2|^2 |\vec{k}|^4,
\end{equation}

\noindent where
\[
Q_2 =\, \frac{1}{2|\vec{k}|^4} \left(3\beta - \alpha|\vec{k}|^2\right).
\]

\noindent The scalar functions \( \alpha \) and \( \beta \) are given by:
\begin{equation}
\begin{aligned}
\alpha &= -gA\, \frac{f^*}{\sqrt{3}m_\pi} \int\, \frac{d^3q}{(2\pi)^3}\, \frac{2M_N}{2E_N(\vec{q}\,)\, 2\omega_{K^{*-}}(\vec{q}\,)\, 2\omega_{\bar{K}^0}(\vec{q}+\vec{k})} \, \frac{1}{P^0 - E_N(\vec{q}\,) - \omega_{K^{*-}}(\vec{q}\,) + i\frac{\Gamma_{K^{*-}}}{2}} \\
&\quad \times\, \frac{1}{P^0 - E_N(\vec{q}\,) - k^0 - \omega_{\bar{K}^0}(\vec{q}+\vec{k}) + i\epsilon} \, \frac{\Lambda^2}{\Lambda^2 + |\vec{q}~^{*2}_{\mathrm{cm}}|}  (2\vec{k} + \vec{q})(a\vec{k} + \vec{q}),
\end{aligned}
\end{equation}

\begin{equation}
\begin{aligned}
\beta &= -gA\, \frac{f^*}{\sqrt{3}m_\pi} \int\, \frac{d^3q}{(2\pi)^3}\, \frac{2M_N}{2E_N(\vec{q}\,)\, 2\omega_{K^{*-}}(\vec{q}\,)\, 2\omega_{\bar{K}^0}(\vec{q}+\vec{k})} \, \frac{1}{P^0 - E_N(\vec{q}\,) - \omega_{K^{*-}}(\vec{q}\,) + i\frac{\Gamma_{K^{*-}}}{2}} \\
&\quad \times\, \frac{1}{P^0 - E_N(\vec{q}\,) - k^0 - \omega_{\bar{K}^0}(\vec{q}+\vec{k}) + i\epsilon} \, \frac{\Lambda^2}{\Lambda^2 + |\vec{q}~^{*2}_{\mathrm{cm}}|}  [(2\vec{k}+\vec{q}) \cdot \vec{k}] \cdot [(a\vec{k}+\vec{q}) \cdot \vec{k}],
\end{aligned}
\end{equation}
where we have  introduced a common form factor $\Lambda^2/(\Lambda^2+\vec{q}~^2)$   with $\Lambda$ typically of the order of 1 GeV, which we adopt here.
\subsection{\large\bfseries Triangle Singularity Condition}
The triangle singularity appears when the internal particles in the loop can simultaneously go on-shell and are classically allowed to propagate. The condition for a triangle singularity is derived from the Coleman-Norton theorem \cite{Coleman:1965xm}, and for this process, it leads to a peak in the $\pi^+ \Lambda$ invariant mass distribution around the $\Sigma^*(1430)$ mass.
\subsubsection{The partial width of $\Sigma^*(1430)$}
For $\Lambda_c^+\longrightarrow \pi^+\pi^-R$ ($R\equiv\Sigma^*(1430)$) we have

\begin{equation}
\begin{aligned}
\frac{d\Gamma}{dM_{\mathrm{inv}}(\pi^-R)}=\frac{1}{(2\pi)^3}\frac{2M_{\Lambda_c^+}2M_{\Lambda}}{4M^2_{\Lambda_c^+}}
p_{\pi^+}\tilde{p}_{\pi^-}\bar{\sum}\sum|t_{\mathrm{TS}}|^2,\label{fomula:1}
\end{aligned}
\end{equation}
with
\begin{equation}
\begin{aligned}
p_{\pi^+}=\frac{\lambda^{\frac{1}{2}}(M^2_{\Lambda^+_c}, m^2_{\pi^+}, M^2_{\mathrm{inv}}(\pi^-R))}{2M_{\Lambda^+_c}},\qquad\qquad
\tilde{p}_{\pi^-}=\frac{\lambda^{\frac{1}{2}}(M^2_{\mathrm{inv}}(\pi^-R), 
m^2_{\pi^-}, M^2_R)}{2M_{\mathrm{inv}}(\pi^-R)}.\label{fomula:1}
\end{aligned}
\end{equation}
Next we give the step to evaluate the mass distribution with to the explicit decay of $\Sigma^*(1430)$ to $\pi^+\Lambda$. As we saw in {Eq.~(\ref{eq:triangle_amplitude})}, to evaluate the loop diagram with $\pi^+\Lambda$ at the end we needed the amplitude $t_{\bar{K}^0p \to \pi^+\Lambda}$. It is easy to establish the link to this case from the coalescence production. One can write \cite{Dias:2018qhp}, $g^2_{\Sigma^{*}(1430), \bar{K}^0p}=-\frac{1}{\pi}\int \mathrm{Im}t_{\bar{K}^0p,~ \bar{K}^0p}(M_\mathrm{inv})dM_\mathrm{inv}$ and then we obtain
\begin{equation}\label{dg1430}
\begin{aligned}
\frac{d\Gamma}{dM_{\mathrm{inv}}(\pi^+ \Lambda)dM_{\mathrm{inv}}(\pi^-R)}=-\frac{1}{\pi}\mathrm{Im}(t_{\bar{K}^0p,\bar{K}^0p})\frac{\Gamma_{\pi^+\Lambda}}{\Gamma_{\mathrm{tot}}}\frac{1}{(2\pi)^3}\frac{2M_{\Lambda_c^+}2M_{\Lambda}}{4M^2_{\Lambda_c^+}}
p_{\pi^+}\tilde{p}_{\pi^-}\bar{\sum}\sum|\tilde{t}_{\mathrm{TS}}|^2,
\end{aligned}
\end{equation}
where we multiply the expression by   ${\Gamma_{\pi^+\Lambda}}/{\Gamma_{\mathrm{tot}}}$, since the coalescence accounted for all   decay channels. We then write     
\begin{equation}
\begin{aligned}
\frac{\Gamma_{\pi^+\Lambda}}{\Gamma_{\mathrm{tot}}}=\frac{\Gamma_{\pi^+\Lambda}}{\Gamma_{\pi^+\Lambda}+\Gamma_{\pi^+\Sigma^0}+\Gamma_{\pi^0\Sigma^+}}=\frac{g^2_{\pi^+\Lambda}\tilde{q}_\Lambda M_\Lambda}{g^2_{\pi^+\Lambda}\tilde{q}_\Lambda M_\Lambda+g^2_{\pi^+\Sigma^0}\tilde{q}_{\Sigma^0} M_{\Sigma^0}+g^2_{\pi^0\Sigma^+}\tilde{q}_{\Sigma^+} M_{\Sigma^+}},\label{fomula:1}
\end{aligned}
\end{equation}
where $\tilde{q}_i$ are the momenta of the $\pi$ in the decay to each channel, and
since $t_{ij}=\frac{g_ig_j}{\sqrt{s}-M_R+\frac{\Gamma_R}{2}}$, we have
\begin{equation}\label{branchgamma}
\begin{aligned}
\frac{\Gamma_{\pi^+\Lambda}}{\Gamma_{\mathrm{tot}}}
&=\frac{|t_{\pi^+\Lambda \to \pi^+\Lambda}|^2\tilde{q}_\Lambda M_\Lambda}{|t_{\pi^+\Lambda \to \pi^+\Lambda}|^2\tilde{q}_\Lambda M_\Lambda+(|t_{\pi^+\Lambda \to \pi^+\Sigma^0}|^2+|t_{\pi^+\Lambda \to \pi^0\Sigma^+}|^2)\tilde{q}_\Sigma M_\Sigma},
\end{aligned}
\end{equation}
with
\begin{equation}
\begin{aligned}
\tilde{q}_\Lambda=\frac{\lambda^{\frac{1}{2}}(M^2_{\mathrm{inv}}(\pi^+\Lambda),m^2_{\pi^+},M^2_\Lambda)}{2M_{\mathrm{inv}}(\pi^+\Lambda)},\\
\tilde{q}_{\Sigma}=\frac{\lambda^{\frac{1}{2}}(M^2_{\mathrm{inv}}(\pi^+\Lambda),m^2_{\pi},M^2_{\Sigma})}{2M_{\mathrm{inv}}(\pi^+\Lambda)}.
\end{aligned}
\end{equation}
We need to calculate $t_{\bar{K}^0p, \bar{K}^0p}$, and the other amplitudes in Eq.~(\ref{branchgamma}) for what we use the chiral unitary approach for meson-baryon scattering amplitudes \cite{Oset:1997it,Hyodo:2011ur}. We extract them by solving the Bethe-Salpeter equation in coupled-channels
\begin{equation}
\begin{aligned}
T=[1-VG]^{-1}V,\label{fomula:1}
\end{aligned}
\end{equation}
where the interaction of the coupled channels is given by
\begin{equation}
\begin{aligned}
V_{ij}=-\frac{1}{4f^2}C_{ij}(k^0_i+k^{'0}_j),\label{fomula:1}
\end{aligned}
\end{equation}
where $k^0_i=\frac{s+m^2_i-M^2_i}{2\sqrt{s}} $, $k^{'0}_j=\frac{s+m^{'2}_j-M^{'2}_j}{2\sqrt{s}} $, with  $m_i,~M_i$  the masses of the initial meson, baryon in the initial state and  $m'_i,~M'_i$  for the final state, with  $s\equiv M_\text{inv}^2(\pi^+\Lambda)$ , with  $C_{ij}$  given by~\cite{Li:2024tvo}, which we reproduce in Table~\ref{table:cij}.
\begin{table}[H]
\centering
\caption{The values of $C_{i j}$ coefficients of different channels in the charge basis.}
\label{table:cij}
\setlength{\tabcolsep}{28pt}
\begin{tabular}{lccccc}
\hline \hline$C_{i j}$ & $\bar{K}^0 p$ & $\pi^{+} \Sigma^0$ & $\pi^0 \Sigma^{+}$ & $\pi^{+} \Lambda$ & $\eta \Sigma^{+}$ \\
\hline $\bar{K}^0 p$ & 1 & $\frac{1}{\sqrt{2}}$ & $-\frac{1}{\sqrt{2}}$ & $\sqrt{\frac{3}{2}}$ & $\sqrt{\frac{3}{2}}$ \\
$\pi^{+} \Sigma^0$ & & 0 & -2 & 0 & 0 \\
$\pi^0 \Sigma^{+}$ & & & 0 & 0 & 0 \\
$\pi^{+} \Lambda$ & & & & 0 & 0 \\
$\eta \Sigma^{+}$ & & & & & 0 \\
\hline \hline
\end{tabular}
\end{table}
The meson-baryon loop function  is regularized with a cutoff,
\begin{equation}
\begin{aligned}
G_i=2M_i\int_{|\vec{q}|<q_{\mathrm{max}}}\frac{d^3q}{(2\pi)^3}\frac{\omega_1(\vec{q}\,)+\omega_2(\vec{q}\,)}{2\omega_1(\vec{q}\,)\omega_2(\vec{q}\,)}\frac{1}{s-[\omega_1(\vec{q}\,)+\omega_2(\vec{q}\,)]^2+i\epsilon}.\label{fomula:1}
\end{aligned}
\end{equation}
with $q_{\mathrm{max}}=630$ ~MeV~\cite{Li:2024tvo}.
\subsubsection{The partial width of $\Sigma^*(1385)$}
Following the same step as in the former subsection, the partial width is
\begin{equation}
\begin{aligned}
\frac{d\Gamma}{dM_{\mathrm{inv}}(\pi^+\Lambda)dM_{\mathrm{inv}}(\pi^-R)}=-\frac{1}{\pi}\mathrm{Im}(t_{\bar{K}^0p \to \bar{K}^0p})\frac{\Gamma_{\pi^+\Lambda}}{\Gamma_{\mathrm{tot}}}\frac{1}{(2\pi)^3}\frac{2M_{\Lambda_c^+}2M_{\Lambda}}{4M^2_{\Lambda_c^+}}p_{\pi^+}\tilde{p}_{\pi^-}\bar{\sum}\sum|t^{\Sigma^*(1385)}_{\mathrm{TS}}|^2,\label{fomula:1}
\end{aligned}
\end{equation}
with
\begin{equation}
\begin{aligned}
\mathrm{Im}(t_{\bar{K}^0p \to \bar{K}^0p})=\mathrm{Im}(\frac{1}{M_{\mathrm{inv}}(\pi^+\Lambda)-M_{\Sigma^*(1385)}+\frac{\mathrm{i}\Gamma_{\Sigma^*(1385)}}{2}}),\label{fomula:1}
\end{aligned}
\end{equation}
with $\frac{\Gamma_{\pi^+\Lambda}}{\Gamma_{\mathrm{tot}}}=0.87$~\cite{ParticleDataGroup:2024cfk}.

\section{\large\bfseries  Results}
\label{For3}
In this section, we present the numerical results for the decay $\Lambda_c^+ \to \Lambda \pi^+ \pi^+ \pi^-$. This decay is initiated by the weak $c \to s$ transition at the quark level, followed by hadronization and strong final-state interactions. Special attention is given to the dynamical generation of the $\Sigma^*(1430)$ resonance and the appearance of triangle singularities in the decay amplitude.

\subsection{\large\bfseries Evidence of a narrow $\Sigma^*(1430)$ resonance in the $\pi^+ \Lambda$ invariant mass}
\label{For3_2}

\begin{figure}[H]
  \centering
  \includegraphics[width=0.4\textwidth]{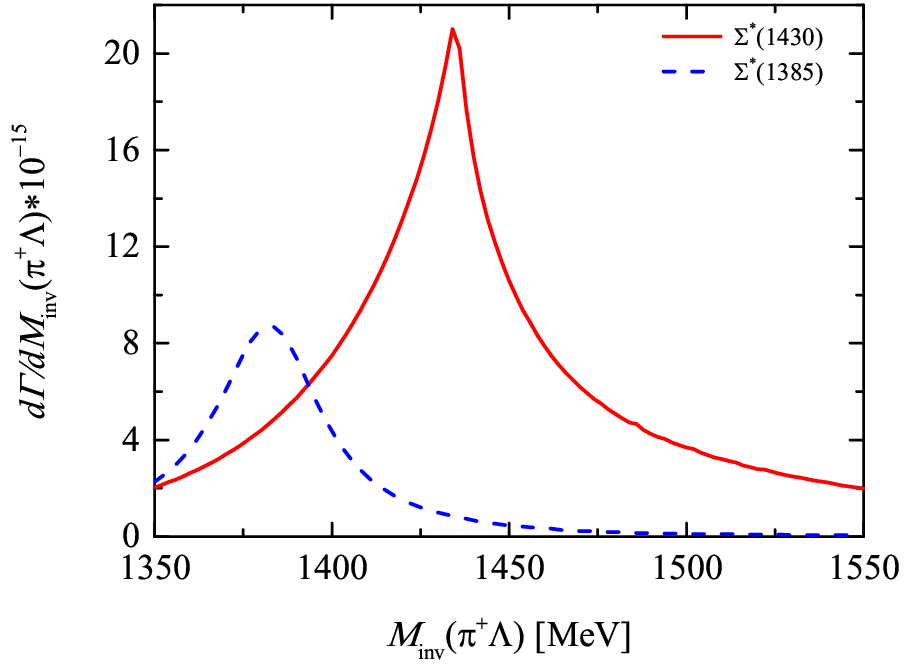}
  \caption{Invariant $\pi^+ \Lambda$ mass distribution in the decay $\Lambda_c^+ \to  \pi^+ \pi^-\Sigma^*(1430)~(\Sigma^*(1430)\to\pi^+\Lambda)$. We also show the contribution of $\Sigma^*(1385)$ through the triangle mechanism of Fig.~\ref{fig:1385new}.
}  \label{Fig:MpiL}
\end{figure}

In Fig.~\ref{Fig:MpiL}, we present the invariant mass distribution $d\Gamma/dM_{\pi^+ \Lambda}$ from Eq.~(\ref{dg1430}), where we have integrated over $M_\text{inv}(\pi^-R)$. A pronounced and very narrow peak is seen near $M_{\pi^+ \Lambda} \simeq 1434$ MeV. This sharp structure is interpreted as a signal of the dynamically generated $\Sigma^*(1430)$ resonance.

The width of this peak is about 25 MeV. From the Belle experiment one finds that $\Gamma_{\Lambda\pi^+}=11.5\pm 2.8\pm 5.3$~MeV, $\Gamma_{\Lambda\pi^-}=33.0\pm 7.5\pm 23.6$~MeV. Hence what we find is in line with the experimental findings. The shape looks like a cusp, which can correspond to a barely bound state~\cite{Li:2024tvo} or barely unbound~\cite{Roca:2013cca} (virtual state). The situation is similar to the case of the $a_0(980)$ resonance in the $\pi \eta$ channel, which lately is getting support as a barely unbound, virtual state~\cite{Super-Kamiokande:2023yqq,Song:2025ofe,ATLAS:2013fep,Li:2024uwu}.

\begin{figure}[H]
    \centering
    \includegraphics[width=0.65\textwidth]{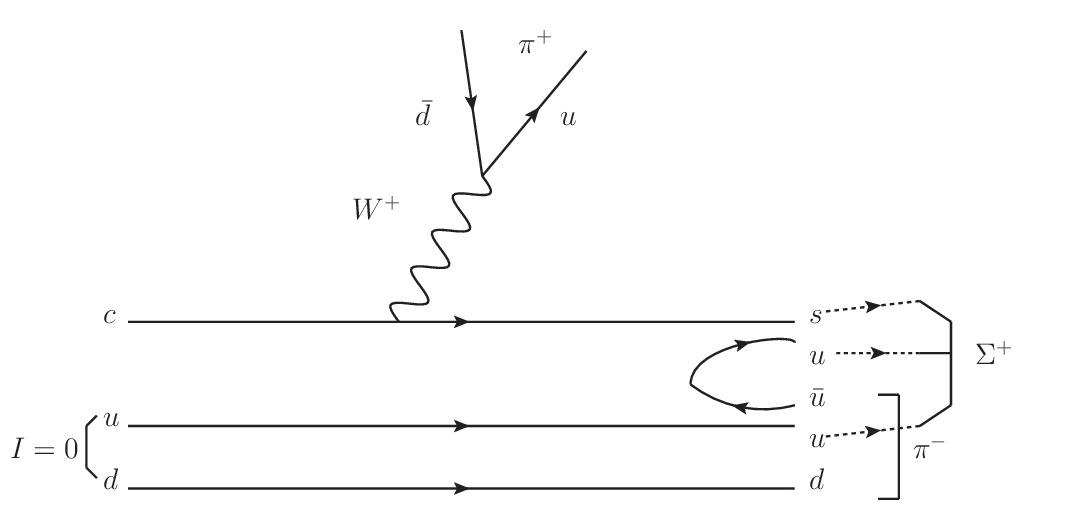}
    \caption{Possible $\Sigma^*(1385)$ weak production in $\Lambda_c^+ \to \pi^+ \pi^- \Sigma^{+}(1385)$.}
    \label{fig:feyndiag_quark2}
\end{figure}

In {Fig.~\ref{Fig:MpiL}} we also show the contribution of the $\Sigma^*(1385)$ excitation via the triangle diagram. As we see, the strength of the mass distribution is small compared to that of the $\Sigma^*(1430)$, which we anticipated because the triangle mechanism does not develop a triangle singularity in this case. We also estimated uncertainties in this term by using  $\Lambda=1.2$ GeV  rather than 1 GeV, and the strength increases by 15\,\%. The large strength of $\Sigma^*(1385)$ excitation {seen} in the Belle experiment is  in favor of the $\Sigma^*(1385)$ being an ordinary 3 $q$ state. A possible topology {for} its  weak production is shown in {Fig.~\ref{fig:feyndiag_quark2}}. However,
\noindent unlike the case of $\Sigma^*(1430)$ production, where we can evaluate the strength of the mass distribution in absolute terms, here we {would} need some free parameters like in {\cite{He:2025vij}} and {the result} would not be a genuine prediction.

\subsection{\large\bfseries Invariant mass distribution of the $\pi^-\Sigma^*(1430)$  system}
\label{For3_1}
\begin{figure}[H]
  \centering
  \includegraphics[width=0.4\textwidth]{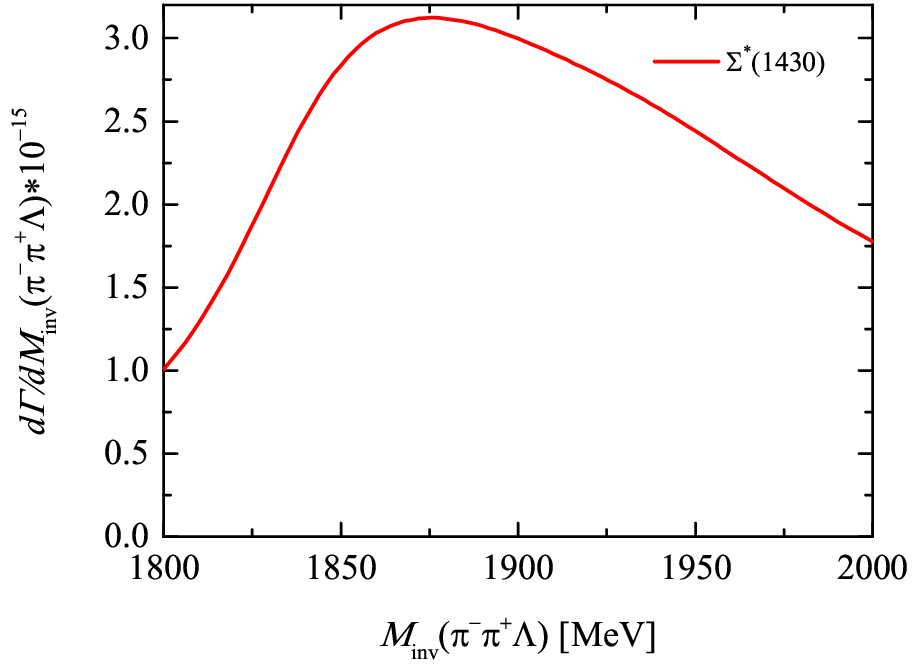}
  \caption{Invariant mass distribution of the $ \pi^- \Sigma^*(1430)~(\Sigma^*(1430)\to\pi^+\Lambda)$ system in the decay $\Lambda_c^+ \to \Lambda \pi^+ \pi^+ \pi^-$. A clear bump appears around 1875 MeV, interpreted as a consequence of a triangle singularity. }
  \label{Fig:M123}
\end{figure}

As we have discussed earlier, the  $\pi^-\Sigma^*(1430)$     mass distribution should develop a triangle singularity. 
In Fig.~\ref{Fig:M123}, we show the invariant mass distribution $d\Gamma/dM_{\pi^- \Sigma^*(1430)~(\Sigma^*(1430)\to\pi^+\Lambda)}$, which corresponds to the $\pi^+ \pi^- \Lambda$ final state. This distribution is obtained by integrating  Eq.~(\ref{dg1430}) over $M_\text{inv}({\pi^+\Lambda})$   coming from the  $\Sigma^*(1430)$ decay. For this we integrate over $M_\text{inv}({\pi^+\Lambda})$   from 1375 MeV to 1500 MeV which covers the whole strength of the   
$\Sigma^*(1430)$  resonance (see Fig.~\ref{Fig:MpiL}). A noticeable enhancement (bump) is clearly observed around $M_{\pi^+ \pi^- \Lambda} \simeq 1875$ MeV.

This enhancement is  attributed to the triangle singularity mechanism, as discussed in Sec.~\ref{sec:formalism} and illustrated by the diagram in Fig.~\ref{fig:feyndiag}. The singularity arises when intermediate particles ($K^{*-}$, $p$, and $\bar K^0$) can simultaneously be on-shell and 
the mechanism can proceed at the classical level, and leads to a sharp increase in the decay amplitude at a specific kinematic point. It would be most instructive to test this experimentally to find out the origin of the $\Sigma^*(1430)$  excitation in this experiment.

\subsection{\large\bfseries Branching ratio estimation}
\label{For3_3}

We estimate the branching ratio for the decay $\Lambda_c^+ \to \pi^+ \pi^-\Sigma^*(1430)~(\Sigma^*(1430)\to\pi^+\Lambda)$ by integrating the invariant mass distributions shown in Fig.~\ref{Fig:MpiL} between 1375 MeV of the 1500 MeV. By considering the $\Lambda_c^+$  width of $3.25\times10^{-9}$~MeV, the resulting value is
\begin{equation}
\text{Br}[\Lambda_c^+ \to \pi^+ \pi^-\Sigma^*(1430)~(\Sigma^*(1430)\to\pi^+\Lambda)] \approx 3.5 \times 10^{-4}.
\end{equation}
This indicates that the decay has a sizable branching fraction, which is within the reach of current experimental measurements as the Belle experiment has shown, but this value has not yet been determined experimentally. The obtained result provides a useful prediction that can be tested in future experiments at facilities such as BESIII, Belle II, or LHCb.

\section{\large\bfseries Conclusion}
\label{sec:conclusion}

In this work, we have studied the decay $\Lambda_c^+ \to \pi^+ \pi^-\Sigma^*(1430)~(\Sigma^*(1430)\to\pi^+\Lambda)$ with a focus on the production mechanism and nature of the $\Sigma^*(1430)$ resonance. Our approach is based on the chiral unitary model, where the $\Sigma^*(1430)$ emerges as a dynamically generated state from meson-baryon interactions in coupled channels. We demonstrated that the $\Sigma^*(1430)$ production can be significantly enhanced through a triangle singularity mechanism involving intermediate $K^{*-}$, $p$, and $\bar K^0$ states. The distributions show a sharp peak in the $\pi^+ \Lambda$ invariant mass distribution near 1434 MeV, which matches well with the structure observed in the Belle experiment.

In addition to identifying the $\Sigma^*(1430)$ resonance, we also predicted a clear bump in the $\pi^-\Sigma^*(1430)$ invariant mass distribution around 1875 MeV, consistent with the kinematic conditions for the triangle singularity. Furthermore, we provided an estimate of the branching ratio for this decay mode, obtaining a value of approximately $3.5 \times 10^{-4}$, which would be most welcome to be contrasted experimentally.

Our results are a consequence of our interpretation of the $\Sigma^*(1430)$ as a non-conventional resonance generated dynamically by final-state interactions, rather than a typical three-quark excitation. The triangle singularity mechanism plays a key role in enhancing the signal of this otherwise elusive state. Future experimental analyses, particularly those that revisit the $\pi^+\Lambda$ and $\pi^+ \pi^- \Lambda$ mass distributions with improved resolution, could offer further evidence for the predictions made in this study and provide deeper insights into the nature of the $\Sigma^*(1430)$.

\section*{ACKNOWLEDGMENTS}
This work is partly supported by the National Natural Science
Foundation of China under Grants  No. 12405089 and No. 12247108 and
the China Postdoctoral Science Foundation under Grant
No. 2022M720360 and No. 2022M720359.  Yi-Yao Li is supported in part by the Guangdong Provincial international exchange program for outstanding young talents of scientific research in 2024. 
 This work is partly supported by the National Natural Science Foundation of China (NSFC) under Grants No. 12365019 and No. 11975083, and by the Central Government Guidance Funds for Local Scientific and Technological Development, China (No. Guike ZY22096024).
 This work is also partly supported by the Spanish Ministerio de Economia y Competitividad (MINECO) and European FEDER
  funds under Contracts No. PID2020-112777GB-I00.
Raquel Molina acknowledges support from
the ESGENT program (ESGENT/018/2024) and the PROMETEU program (CIPROM/2023/59), of the Generalitat Valenciana, and also
from the Spanish Ministerio de Economia y Competitividad and
European Union (NextGenerationEU/PRTR) by the grant
(CNS2022-13614. This research partially supported by grant PID2023-147458NB-C21 funded by MCIN/AEI/ 10.13039/501100011033 and by the European Union.

\bibliography{refs.bib} 
\end{document}